\documentclass[12pt,fleqn]{article}
\usepackage{psfig}
\begin{document}
\begin{center}
{\Large{\bf RADIATIVE PRODUCTION OF THE $\Lambda(1405)$}} 
\end{center}
\begin{center}
{\Large{\bf RESONANCE
IN $K^-$ COLLISIONS}} 
\end{center}
\begin{center}
{\Large{\bf ON PROTONS AND NUCLEI}}
\end{center}

\vspace{1cm}

\begin{center}
{\large{ J.C. Nacher$^ {1,2}$, E. Oset$^ {1,2}$, H. Toki$^1$ and A. Ramos$^3$}}
\end{center}

\vspace{0.4cm}

\begin{center}
{\it $^1$ Research Center for Nuclear Physics (RCNP), Osaka University,}
{\it Ibaraki, Osaka 567-0047, Japan.}
\end{center}
\begin{center}
{\it $^2$  Departamento de F\'{\i}sica Te\'orica and IFIC, 
Centro Mixto Universidad de Valencia-CSIC
46100 Burjassot (Valencia), Spain.}
\end{center}
\begin{center}
{\it $^3$ Departament d'Estructura i Constituents de la Materia, Universitat de
Barcelona,}
{\it Diagonal 647, 08028 Barcelona, Spain.}
\end{center}
\vspace{2.2cm}

\begin{abstract}
{\small{We have carried a theoretical study of the $K^- p\rightarrow M B
\gamma$ reaction with $M B = K^-p, \bar{K}^0 n, \pi^- \Sigma^+, 
\pi^+ \Sigma^-, \pi^0 \Sigma^0, \pi^0 \Lambda$, for $K^-$ lab. momenta 
between 200 and 500 MeV/c, using a chiral unitary approach for the strong
$K^-p$ interaction with its coupled channels. The $\Lambda(1405)$ resonance,
 which is generated dynamically in this approach, shows
 up clearly in the $d\sigma/dM_I$ spectrum, providing new tests for 
 chiral symmetry and the unitary approach, as well as information regarding
 the nature of the resonance. The photon detection alone, summing all channels,
 is shown to reproduce quite accurately the strength and shape of the
 $\Lambda(1405)$ resonance. Analogous reactions in nuclei can provide
 much information on the properties of this resonance in a nuclear medium.}}
 
\end{abstract}
\vspace{2.2cm}
\newpage

The $\Lambda$(1405) resonance, $S_{01}$, I($J^P$)=0($\frac{1}{2}^-$),
appears just below the $K^-p$ threshold and plays an essential role in the
$K^-p$ interaction and coupled channels at low energies. The modification
 of its properties in nuclei influences directly the $K^-$ nucleus interaction,
 a topic which has raised much theoretical interest [1, 2, 3] in order to
 explain the attraction manifested by empirical analyses of $K^-$ atoms [4] and
 which eventually could lead to $K^-$ condensation in neutron stars [5].

 Another attractive feature of the $\Lambda$(1405) resonance is its dynamical
 generation within the context of chiral unitary theories [6, 7, 8] which
 lead to a good description of the low energy $\bar{K} N$ interaction
 and coupled channels. The dynamical generation of the $\Lambda(1405)$
 in the chiral context has larger repercusion than just the description of
 these strong interactions in the strangeness $S=-1$ sector. It also leads
 to define mechanisms for its electromagnetic production or decay. The strength
 of these mechanisms also changes from one reaction to another and as a
 function of the energy. Hence, the investigation of these reactions offers
 a test of chiral symmetry and of the nonperturbative methods used to generate
 the low lying resonances. One such test, which is passed successfully, is
 offered by the $K^- p\rightarrow\Lambda\gamma$, $\Sigma^0\gamma$ reaction, where
  the unitarity in coupled channels enhances the $\Sigma^0\gamma$ production by
  more than one order of magnitude [9]. Photoproduction processes in the
  $S=0$ sector within the chiral schemes have equally proved successful
  [7].
  
  In a recent paper [10] a study of the $\Lambda(1405)$ photoproduction process
  close to threshold was done in the $\gamma p\rightarrow K^+\Lambda(1405)$
   reaction, showing that the $K^+$ detection alone was a good tool to observe
   the resonance. This reaction, implemented in nuclei, was already suggested
    in [1] as a means to observe modifications of the $\Lambda(1405)$
    resonance in a nuclear medium. In [10] was shown that the study of the resonance in nuclei
   required the specific detection of the $\pi\Sigma$ decay channels of the
   $\Lambda(1405)$, the Fermi motion blurring any trace of the resonance
   if only the $K^+$ is detected.
   
   In the present paper we study a crossed channel of the former reaction,
   namely the $K^-p\rightarrow\Lambda(1405)\gamma$ reaction at low $K^-$
   energies. One novelty of the present reaction is that the photon has now
   small energy, of the order of 100 MeV in the energy regime studied here,
   while the photon required in the $\gamma p\rightarrow K^+ \Lambda(1405)$
   reaction had around 1.7 GeV. The smaller energy of the $K^-$ and $\gamma$ in the
   $K^- p\rightarrow\Lambda(1405)\gamma$ reaction makes the effects of Fermi
   motion in nuclei much weaker, such that the detection of the photon still 
   leads to 
   a resonant shape in the cross section. Another novelty is that 
   the reactions are rather different dynamically, with the dominant mechanisms in the first reaction being negligible in the
   second one, and others which could be proved negligible in the first one
   becoming now dominant in the second reaction. Hence, in spite of the
   similarities of the two reactions, the information provided by them as tests
   of the chiral approaches are different and complementary.
   
   We shall follow the chiral approach of [8] for the strong interaction in the
   $S=-1$ sector where the Bethe Salpeter [BS] equation is solved with the
   coupled channels $K^-p, \bar{K}^0 n, \pi^0\Lambda, \pi^0\Sigma^0,
   \eta\Lambda, \eta\Sigma^0, \pi^+\Sigma^-, \pi^-\Sigma^+, K^+\Xi^-,
   K^0\Xi^0$.
   
   The BS equation reads in matrix form:
   
\begin{equation}
T = V + VGT \rightarrow T = [1 - VG]^{-1}V\, ,   
\end{equation}
  {
   with G a diagonal matrix, accounting for a loop with a meson
   and a baryon propagator, with matrix elements
   
\begin{equation}  
G_{l}(s)  =  i\int\frac{d^4 q_L}{(2\pi)^4}\, \frac{M_l}{E_l(\vec{q}_L)}
\, \hspace{0.2cm}\frac{1}{\sqrt{s} - q_L^0 - 
E_l(\vec{q}_L) + i\epsilon}\, \hspace{0.2cm} \frac{1}{q_L^2 - m_l^2 + i\epsilon}
\end{equation}
and V is the transition matrix extracted from the lowest order chiral
Lagrangian, which in the s-wave and usual low energy approximation reads as [8]

\begin{equation}
V_{ij} = - C_{ij}\frac{1}{4f^2}(k^0 + k'^0) ,
\end{equation} 
where $k$, $k'$ are the momenta of the incoming, outgoing mesons, $f$ is the
 pion decay constant ($f_{\pi} = 93$ MeV) and $C_{ij}$ a $10\times 10$ matrix of
 coefficients which are shown in table 1 of [8]. In [8] an average value
 $f=1.15 f_{\pi}$, between the one for pions and kaons is used. This, together
 with the choice of a cut off, $|\vec{q}_L|_{max}$, in the integral of eq. (2),
  provides an effective way of accounting for effects of the higher order
  Lagrangians in this particular sector (see ref. [11] for
  an interpretation within the more general inverse amplitude method). Eq. (1)
  is an algebraic equation, and the loop integral affects only the meson and
  baryon propagators in $G_l$, since it was proved in [8] that V and T
  factorized outside the integral with their on shell values.
  
  The series of diagrams summed up by the BS equation is shown in line (1) of fig
  1. From there we can draw all diagrams in which an external photon is coupled
  in addition, which leads to the rest of diagrams shown in the figure.
  
  Apart from the strong $MB\rightarrow M'B'$ vertices of eq. (3) we need
  now the coupling of the photon to the baryons, the mesons, plus the contact
   term of diagram (2,a) of fig. 1 required by gauge invariance. These vertices
   are standard and after the nonrelativistic reduction of the $\gamma$
   matrices, are given in the Coulomb gauge, $\epsilon^0 = 0$, 
   $\vec{\epsilon}\cdot\vec{q}=0$, with $\vec{q}$ the photon momentum, by

\begin{equation}
a) -it_{M'M\gamma} = 2ieQ_M\vec{k}'\cdot\vec{\epsilon}
\end{equation}
for the coupling of the photon to the mesons, with $e$ electron charge, 
$Q_M$ the charge of the meson, $k'$ the momentum of the outgoing meson
and $\epsilon_{\mu}$ the photon polarization vector.

\begin{equation}
b) -it_{B'B\gamma} = ie(Q_B\frac{\vec{p}+\vec{p}\, '}{2M_B} -i \frac{
\vec{\sigma}\times\vec{q}}{2M_B}\mu_B)\vec{\epsilon} 
\end{equation}
for the coupling of the photon to the baryons, with $Q_B$ the baryon charges,
$\vec{p}$, $\vec{p}\, '$ incoming and outgoing momenta of the baryon and $M_B$,
$\mu_B$ the mass and the magnetic moment of the baryon.

\vspace{1.5cm}
\begin{figure}[h]
\centerline{\protect
\hbox{
\psfig{file=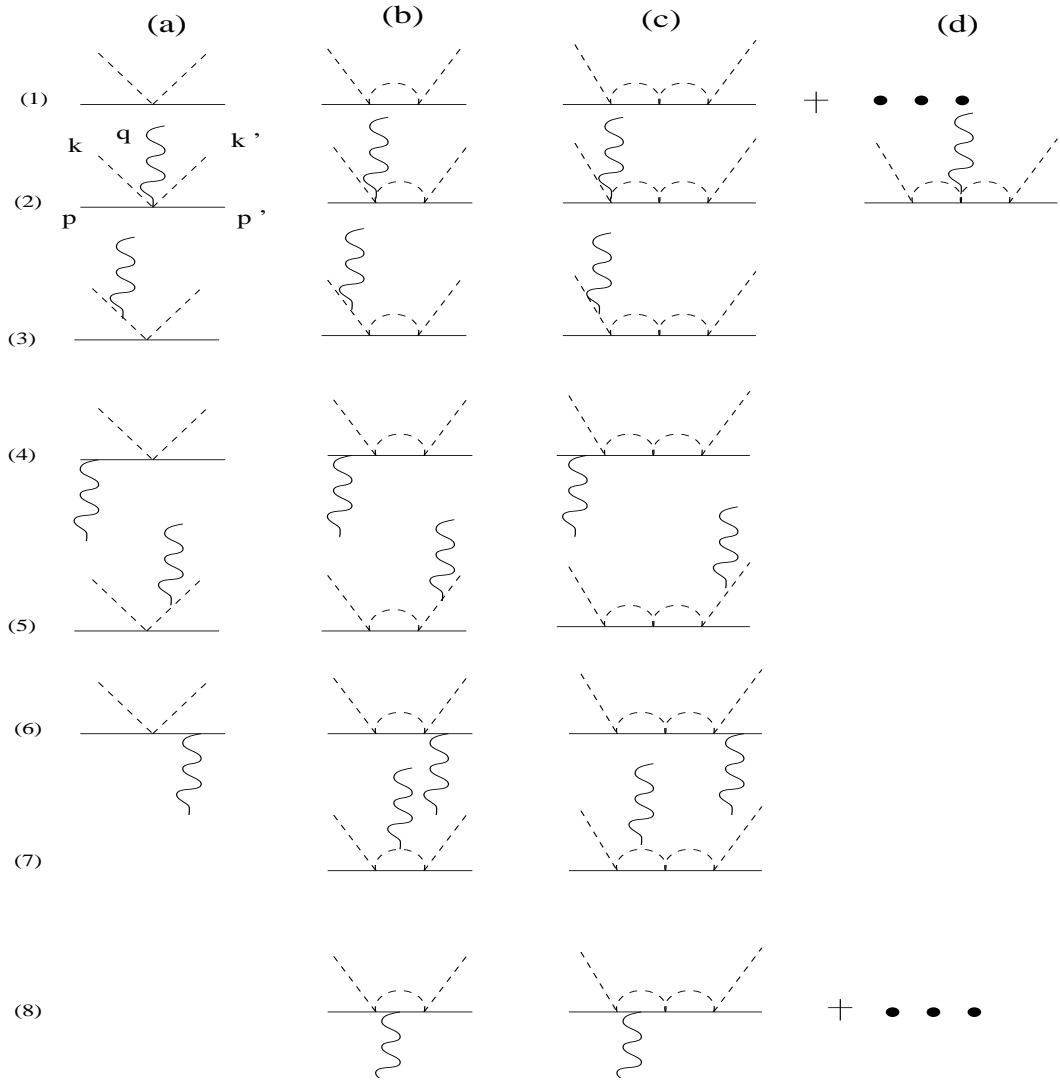,height=14.5cm,width=14.0cm,angle=0}}}
\caption{Feynman diagrams used in the model for the $K^-
p\rightarrow \Lambda(1405)\gamma$ reaction. }
\label{Fig.1}
\end{figure}
 
\begin{equation}
c) -it_{B'M'B M\gamma} = iC_{ij}(Q_i + Q_j)\{\frac{\vec{p}+\vec{p}\, '}{2\bar{M}} 
-i \frac{
\vec{\sigma}\times(\vec{p}-\vec{p}\, ')}{2\bar{M}}\}\vec{\epsilon} 
\end{equation}
for the contact term of diagram (2,a) of fig. 1, with $C_{ij}$ the coefficients
of eq. (3), $i,j$ standing for a $BM$ state, $Q_i$, $Q_j$ the charges of the
mesons, $\bar{M}$ an average mass of the baryons and $\vec{p}$, $\vec{p}\, '$ the
momenta of the incoming and outgoing baryons.

We shall be concerned with $K^-$ with momenta below 500 MeV/c in the
lab frame. In this energy domain it is easy to see that the Bremsstrahlung
 diagrams from mesons and baryons (diagrams (3,a)(4,a)(5,a)(6,a)) are of the same
 order of magnitude and that the contact term (diagram (2,a)) is of order
 $q/2M$ of the corresponding meson Bremsstrahlung diagrams ((3,a) and (5,a)). With CM
 photon momenta $q$ of the order of 150 MeV or below, these terms represent
 corrections below the 8$\%$ level and we shall neglect them. In addition, terms
 like in diagram (2,d), where the photon couples to internal vertices of the
 loops, vanish for parity reasons.
 
 The diagrams of row (7) of fig. 1 where the photon couples to mesons in the
 loops vanish due to the gauge condition $\vec{\epsilon}\cdot\vec{q} = 0$
 and the same happens to the diagrams in row (8) where the photon couples with 
 the dielectric part ($(\vec{p}+\vec{p}\, ')$ term of eq. (5)) to
 the baryons inside the loops. This occurs because in both cases one has
 integrals of the type  $\vec{\epsilon}\int d^3 k\vec{k}\cdot f(k,q)$ and the integral gives a vector proportional
 to $\vec{q}$, the only vector in the integrand which is not integrated when we
 work in the CM frame. The magnetic coupling of the photons in row (8)
 survives.

 Hence, the process is given, within the approximations mentioned, by the
 diagrams of the rows 3, 4, 5, 6 plus the magnetic part in row 8. This situation is opposite to the
 one found in [10] for $\Lambda(1405)$ photoproduction close to threshold, where
 the dominant terms came from the contact term and the Bremsstrahlung diagrams were
 negligible.
 
 If we inspect the series of terms in rows (3) and (4) of fig. 1 we see that
 the strong part of the interaction to the right of the
 electromagnetic vertex involves the sum $V + VGV +\cdots =
 V + VGT$. This is the $T$ matrix from the initial $MB$ state to the
 final $M'B'$ state after losing the energy of the photon, this is, with an
 argument $M_I$, where $M_I$ is the invariant mass of the $M'B'$ state.
 Similarly, in the rows (5) (6) the strong $T$ matrix factorizes before the
 electromagnetic vertex with an argument $\sqrt{s}$, with $s$ the Mandelstam
 variable for the initial $K^- p$ system. In the diagrams of row (8) we have
 a loop with one meson and two baryons. The strong interaction to the left
 originates $t(\sqrt{s})$ and the one to the right $t(M_I)$. In order to
 evaluate the loop function we note that $q\ll \omega(q_L)$ and neglecting $q$
 in the second baryon propagator we observe that the baryon propagator appears
 squared and hence this loop function can be obtained by differentiation
 of the loop of eq. (2) with respect to $\sqrt{s}$, which is what we do.
  This term gives a contribution of less than $10\%$ in the cross section and
  this approximation is hence justified.

 Thus, a straighforward calculation, which includes the vertices of eqs.(4, 5),
 plus the meson and baryon propagators, allows one to write the whole amplitude
 for the process in terms of an electric and a magnetic part as:
 
\begin{equation}
t_{ij}^{(\gamma)}=t_{ij}^{(1E)} \vec{k}\cdot\vec{\epsilon} +
t_{ij}^{(2E)} \vec{k'}\cdot\vec{\epsilon} +
it^{(M)}(\vec{\sigma}\times\vec{q})\cdot\vec{\epsilon}\, ,
\end{equation}
where

\begin{equation}
\hspace{-0.8cm}
t_{ij}^{(1E)} = e Q_i(\frac{1}{k\cdot q} + \frac{1}{q M_i})t_{ij}(M_I)\, ,
\end{equation}
\begin{equation} 
\hspace{-0.8cm}
t_{ij}^{(2E)}= -e Q_j(\frac{1}{k'\cdot q} + \frac{1}{q M_j})t_{ij}(\sqrt{s})\,
, 
\end{equation}

\begin{equation}
\hspace{-0.8cm}
t^{(M)}= -\frac{e\mu_{Bi}}{2qM_i}t_{ij}(M_I) +
\frac{e\mu_{Bj}}{2qM_j}t_{ij}(\sqrt{s})\, - \sum_l t_{il}(\sqrt{s})
\frac{\partial{G_l(s)}}{\partial{\sqrt{s}}}t_{lj}(M_I)\frac{e\mu_{Bl}}{2M_l}
\end{equation}

The fact that in the amplitudes of eqs. (8-10) the strong $t$ matrix appears
 sometimes with the argument $M_I$ makes it convenient to evaluate the phase
 space for the cross section with $d^3 q/2 q$ in the CM of the $K^-
 p$ system and $d^3 k'/2w(k')$ in the CM of the final meson baryon
 system. Hence we have

\begin{equation}
k\cdot q= k^0 q^0 - kqcos\theta\, ,
\end{equation}
\begin{equation}
k'\cdot q= \tilde{k}'^0\tilde{q}^0 - \tilde{k}'\tilde{q}cos\theta'\, , 
\end{equation}
with $\tilde{k}'$, $\tilde{q}$ the final meson and photon momenta in the final meson
baryon CM frame,
\begin{equation} 
\tilde{q}=\frac{q\sqrt{s}}{M_I};\hspace{0.5cm} \tilde{k}' = \frac{\lambda^{1/2}(M_I^2, m_j^2, M^2_j)}{2M_I}
 ;\hspace{0.5cm} \tilde{k}^0 = \frac{M^2_I + m_j^2 - M_j^2}{2M_I}\, ,
\end{equation}
with $m_j$, $M_j$ the final meson, baryon masses, $\theta$ the angle between 
 $\vec{q}$ and $\vec{k}$ in the $K^- p$ CM frame and $\theta'$ the angle
 between $\vec{k}\, '$ and $\vec{q}$ in the final meson baryon CM frame. By
 choosing an appropiate pair of orthogonal $\epsilon_1, \epsilon_2$ photon
 polarization vectors, also orthogonal to $\vec{q}$, and summing over photon
 and final baryon polarizations plus averaging over the initial proton
 polarization, we 
 finally obtain
  the cross section for the process given by
 ($\sigma$ the
 cross section for each $i, j$ transition)
 
\begin{equation}
\frac{d\sigma}{dM_Id\varphi} = \frac{1}{2\pi}\frac{d\sigma}{dM_I} +
\frac{d\sigma_I}{dM_Id\varphi}cos\varphi\, ,
\end{equation}
with $\varphi$ the azimutal angle formed by the plane containing the $\vec{k}'$
 and $\vec{q}$ vectors and the one containing the $\vec{k}$ and $\vec{q}$
 vectors. The only dependence on the azimutal angle $\varphi$ comes in the
 $cos\varphi$ dependence which accompanies the interference cross section,
 $\sigma_I$, in eq. (14), which means that both $d\sigma/dM_I$ and $d\sigma_I/
 dM_Id\varphi$ do not depend on the angle $\varphi$.

 The expresions for $d\sigma/dM_I$ and $d\sigma_I/dM_Id\varphi$
 are given by

\begin{equation} 
\frac{d\sigma}{dM_I} = 2\pi F\{|t^{(1E)}|^2 k^2 sin^2\theta  + |t^{(2E)}|^2  
\tilde{k}'^2
sin^2\theta' + 2|t_M|^2 q^2\}\, ,
\end{equation}
\begin{equation}
\frac{d\sigma_I}{dM_Id\varphi} = F\{ -2 Re (t^{(1E)}t^{(2E)\ast}) k
\tilde{k} 'sin\theta 
sin\theta '\}\, ,
\end{equation}

with F an operator symbolizing
$$ 
F\rightarrow\frac{1}{(2\pi)^4}\frac{1}{4s}\frac{M_iM_j}{\lambda^{1/2}(s, M_i^2,
m_i^2)}\frac{1}{M_I}(s - M_I^2)\lambda^{1/2}(M_I^2, M_j^2, m_j^2)
$$

\begin{equation}
\times\frac{1}{2}\int_{-1}^{1}dcos\theta\frac{1}{2}\int_{-1}^{1}dcos\theta'\, .
\end{equation}

 We show the results in figs. 2 and 3. In fig. 2 we can see the results for the
  cross sections in the $K^- p\rightarrow\pi^-\Sigma^+\gamma, 
 \pi^+\Sigma^-\gamma, \pi^0\Sigma^0\gamma, \pi^0\Lambda\gamma$, $K^- p \gamma$ 
 channels. The cross section for $K^-p\rightarrow\bar{K}^0 n\gamma$ is very
  small, around 0.1 mb\, GeV$^{-1}$ in the range $1.44 - 1.52$ GeV, and is not
  plotted in the figure. The $\Lambda(1405)$ peak appears clearly in the $\pi\Sigma$
 spectrum. It is interesting to notice the difference between the cross
 sections
 for the different $\pi\Sigma$ channels. The origin of this is the same
 one discussed in [10] due to the different isospin combinations of the
 three charged states and the crossed products of the I=1, I=0 amplitudes
 which appear in the cross section. The $\pi^0\Sigma^0$ has no I=1 component
 and since the I=2 component is negligible, the $\pi^0\Sigma^0$ distribution is very
 similar to the I=0 $\Lambda(1405)$ distribution. On the other hand, 
 the I=1, I=0 crossed products cancel in the sum
 of the $\pi^+\Sigma^-$, $\pi^-\Sigma^+$ distributions and the remaining  
 sum of the squares of the I=0 and I=1
 amplitudes is largely dominated by the I=0 contribution. Hence,
 both
 the $\pi^0 \Sigma^0$ distribution or the sum of the three $\pi\Sigma$ channels
 have approximately the shape of the resonance.  The I=0 contribution alone,
 coming from the excitation of the $\Lambda(1405)$, can be obtained using a
 combination of the three $\pi\Sigma$ amplitudes. Given the isospin
 decomposition of the $\pi\Sigma$ states, the combination that gives the $I=0$
 component is given by:

\begin{equation}
(t_{K^- p\rightarrow\pi^-\Sigma^+} + t_{K^- p\rightarrow\pi^+\Sigma^-} + 
 t_{K^- p\rightarrow\pi^0\Sigma^0})/\sqrt{3}\, ,
\end{equation} 
Hence, the $I=0$ cross section can be obtained using eqs. (8-10) substituting
$t_{ij}$ by the combination of eq. (18).

 We show the results for the pure I=0 excitation
 in fig. 2 which looks very similar to the total strengh around the
 $\Lambda(1405)$ peak.
 
\vspace{1cm}
\begin{figure}[h]
\centerline{\protect
\hbox{
\psfig{file=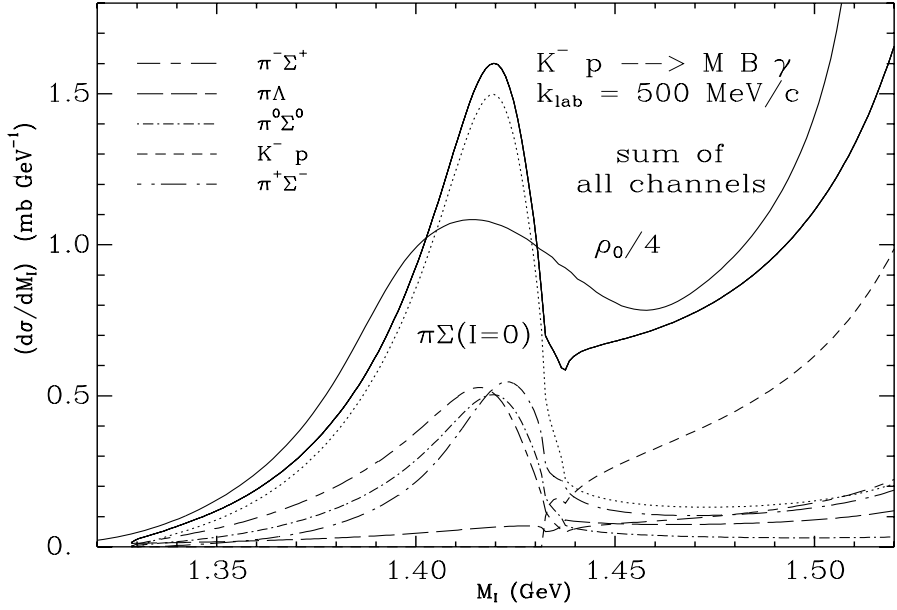,height=8cm,width=12.5cm,angle=-90}}}
\caption{Mass distribution for the different channels, eq. (15). The solid line
with the resonance  shape is the sum of cross sections for all channels.
Dotted line: pure $I=0$ contribution from the $\Sigma\pi$ channels.
The effects of the Fermi motion with ($\rho=\rho_0/4$) is shown with solid
line. The labels for the other lines are shown in the figure.}
\end{figure}
\vspace{0.5cm}

 Below the $K^- p$ threshold there is some strength for I=1,
 $\pi^0\Lambda$ excitation, which is also very small as shown in the figure. As
 a consequence of that, the sum of all channels in the $\Lambda(1405)$
 region, which requires exclusively the detection of the photon, has
 appoximately the $\Lambda(1405)$ shape and strength.
 
 It is interesting to observe the fast rise of the cross section in the
 $K^-p\rightarrow K^-p\gamma$ channel, showing the Bremsstrahlung infrarred
 divergence at large $M_I$ (small photon momentum). The other channels also would show the infrarred divergence at higher
 energies,
  when the photon momentum goes to zero. The relative larger 
  weight of the $K^- p\rightarrow K^- p\gamma$ reaction at these
  energies, with
  respect to the other ones, is a reflection of the fact that the
  $K^-p\rightarrow K^-p$ cross section at values of $M_I$ or $\sqrt{s}$ of the
  order of 1500 is much bigger than the other $K^-p\rightarrow M'B'$ cross
  sections. 
 
 In fig. 3 we show $d\sigma_{I}/dM_Id\varphi$. This magnitude is interesting
 since it involves the  $Re(t^{(1E)}t^{(2E)\ast})$ and hence
 $Re(t(M_I)t(\sqrt{s})^\ast)$. Since $t(\sqrt{s})$ is fixed, the interference cross section as a function of $M_I$ shows variations of $t(M_I)$ in a different
 combination than the one appearing in $d\sigma/dM_I$ of fig. 2, adding
 complementary information on the strong scattering amplitudes, and also
 further testing the chiral approach used, which gives a specific weight to the
 different terms. The approximate different signs in the $\pi^-\Sigma^+\gamma$
 and $\pi^+\Sigma^-\gamma$ channels observed in the figure reflect the product $Q_iQ_j$ which appears
 in the product of $t_{ij}^{(1E)}$$t_{ij}^{(2E)}$. This also tells us that the
 channels with neutral particles in the final state, $\bar{K}^0n$,
 $\pi^0\Lambda$, $\pi^0\Sigma^0$, do not have any contribution to 
 $d\sigma_I/dM_I$. This is of course another test of the approach which could
 be tested experimentally.

The results obtained here involve the use of the $K^- p\rightarrow M B$
interaction at higher $K^-$ momenta than tested in [8], where $K^-$ lab
was below 200 MeV/c. The experimental cross sections extrapolate smoothly
at higher energies in the range used here and the model of [8] still provides
a fair description of the data [12,13,14]. Note, 
moreover, that the peak of $d\sigma/dM_I$
around the $\Lambda(1405)$ resonance is dominated by $t_{ij}(M_I)$ with 
$M_I\simeq 1400$ MeV where the model of [8] proved to be very accurate. 
 
\vspace{1cm}
\begin{figure}[h]
\centerline{\protect
\hbox{
\psfig{file=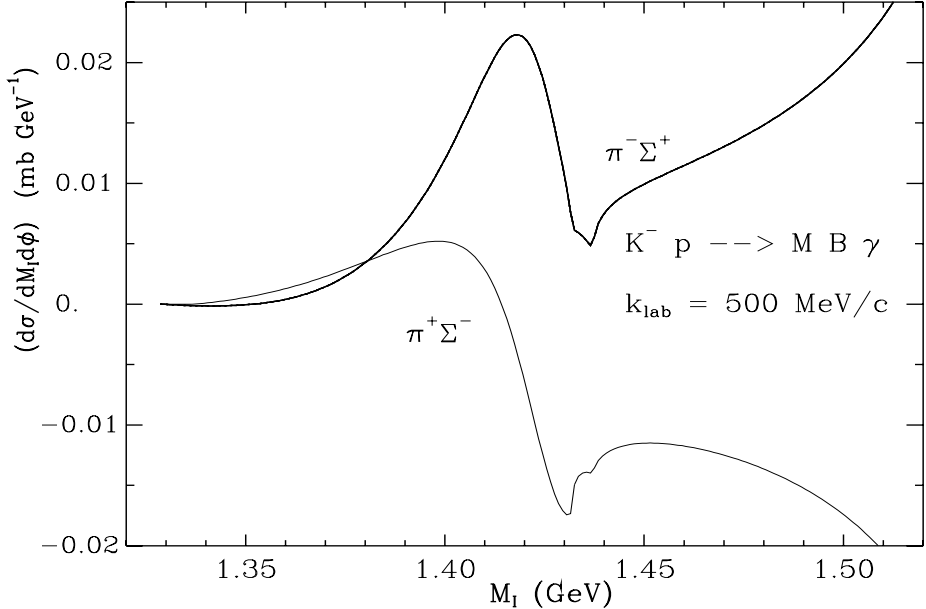,height=8cm,width=12.5cm,angle=-90}}}
\caption{Double differential cross section $d\sigma_I/dM_Id\varphi$ 
for $\Sigma^+\pi^-$ and $\Sigma^-\pi^+$.}
\end{figure}
\vspace{0.5cm}

 We have also performed calculations at lower energies of the $K^-$ and the
 results are qualitatively very similar. There is, however, a lower limit
 to this experiment. Indeed, when the Bremsstrahlung tail of the $K^-p\rightarrow
 K^- p\gamma$ reaction overlaps with the $\Lambda(1405)$ signal 
 all valuable information obtained
 by detecting the photon alone is lost. 
 We have checked that this occurs for $K^-$ lab.
  momenta below 200 MeV/c. However, if one detects the $\pi\Sigma$ final
  particles in coincidence, due to the relatively smaller weight of the cross
  section at high energies, one can go to smaller $K^-$ momenta before the
  Bremsstrahlung tail also blurs the image of the $\Lambda(1405)$ resonance. We have
  checked that this happens below $K^-$ lab. momenta of 150 MeV/c.
  
  The evaluation of the corresponding reactions in nuclei would require to
  account for the distortion of the incoming $K^-$ waves. In this case if one
  detects only the photon one has a distribution of invariant masses due to
  Fermi motion since now $M_I^2 = (k + p_N - q)^2$ and $p_N$ runs over all nucleon
  momenta of the
  occupied states. We have folded our $d\sigma/dM_I$ results with the
  distribution of $M_I$ coming from a Fermi sea of nucleons and show the
  results in fig. 2 for $\rho = \rho_0/4$, a likely
  effective density for this reaction, taking into account $K^-$ distortion.
  We can see a widening of the $\Lambda(1405)$ distribution, with the shape
  only moderately changed
  , such that other effects from genuine changes
  of the $\Lambda(1405)$ properties in the medium, predicted to be quite
  drastic [1, 2, 3] could in principle be visible. Certainly the detailed
  measurement of the final meson baryon in coincidence with the photon would
  allow a 
  much better determination of the $\Lambda(1405)$ properties than just the
  photon detection, and ultimately these exclusive measurements should also be
  performed.

 The proposed reactions can be easily implemented at present facilities like
 KEK or Brookhaven. In Brookhaven some data from recent $K^- p$
 experiment with detection of photons in the final state are in the process of
  analysis [15]. The present results should encourage the detailed
 analysis of the particular channels discussed here.

\vspace{3cm}

{\bf Acknowledgements.}
 We are grateful to the COE Professorship program of Monbusho, which enabled E. O. to
stay at RCNP to perform the present work. One of us, J.C. Nacher would like to acknowledge the hospitality of the RCNP of the Osaka University where
	this work was done and support from the Ministerio de Educacion y Cultura. This work is partly supported
	by DGICYT contract number PB96-0753 and PB95-1249.

\end{document}